\documentclass{amsart}
\usepackage{amsfonts}
\usepackage{amsmath}
\usepackage{amssymb}
\usepackage{graphicx}

\theoremstyle{plain}
\newtheorem{theorem}{Theorem}[section]

\newtheorem{proposition}[theorem]{Proposition}
\theoremstyle{definition}

\theoremstyle{remark}
\newtheorem{remark}[theorem]{Remark}
\numberwithin{equation}{section}

\begin{document}
\title[The vertex solution of a Riemann type hydrodynamical equation]{A
vertex operator representation of solutions to the Gurevich-Zybin
hydrodynamical equation}
\author{Yarema A. Prykarpatsky$^{1,2}$}
\address{$^{1}$The Department of Applied Mathematics at the University,
Krakow 30059, Poland\\
and\\
$^{2}$Department of Differential Equations of the Institute mathematics at
NAS, Kyiv, Ukraine}
\email{yarpry@gmail.com}
\author{Denis Blackmore$^{3}$}
\address{$^{3}$Department of Mathematical Sciences and Center for Applied
Mathematics and Statistics, New Jersey Institute of Technology, Newark, NJ
07102, USA}
\email{deblac@m.njit.edu}
\author{Jolanta Golenia}
\address{$^{4}$The Department of Applied Mathematics at the AGH University
of Science and Technology, Krakow 30059, Poland\\
goljols@tlen.pl}
\author{Anatoliy K. Prykarpatsky$^{5,6}$}
\address{$^{5}$The Department of Mining Geodesy and Environment Engineering
at the AGH University of Science and Technology, Krakow 30059, Poland\\
and\\
$^{6}$Department of Economical Cybernetics at the Ivan Franko Pedagogical
State University, Drohobych, Lviv region, Ukraine \ }
\email{pryk.anat@ua.fm, prykanat@cybergal.com}
\date{present}

\begin{abstract}
An approach based on the spectral and Lie - algebraic techniques for
constructing vertex operator representation for solutions to a Riemann type
hydrodynamical hierarchy is devised. A functional representation \
generating an infinite hierarchy of dispersive Lax type integrable flows \
is obtained.
\end{abstract}

\maketitle

\section{\protect\bigskip Introduction}

Nonlinear hydrodynamic equations are of constant interest still from
classical works by B. Riemann, who had extensively studied them in general
three-dimensional case, having paid special attention to their
one-dimensional spatial reduction, for which he devised the generalized
method of characteristics and Riemann invariants. \ These methods appeared
to be very effective \cite{Wh} in investigating many types of nonlinear
spatially one-dimensional systems of hydrodynamical type and, in particular,
the characteristics method in the form of a "reciprocal" transformation of
variables has been \ used recently in studying a so called Gurevich-Zybin
system \cite{GZ,GZ1} in \cite{Pav} and a Whitham type system in \cite%
{PrPryt,GPPP}. Moreover, this method was further effectively applied to
studying solutions to a generalized \cite{GPPP} (owing to D. Holm and M.
Pavlov) Riemann type hydrodynamical system
\begin{equation}
D_{t}^{N}u=0,\text{ \ \ }D_{t}:=\partial /\partial t+u\partial /\partial x,%
\text{ \ }  \label{Z1}
\end{equation}%
where $N\in \mathbb{Z}_{+}$ and $u\in C^{\infty }(\mathbb{R}^{2};\mathbb{R})$
is a smooth function. Making use of novel methods, devised in \cite%
{PAPP,PoP} and based both on the spectral theory \cite{Nov,PM,MBPS,HPP} and
the differential algebra techniques, the Lax type representations for the
cases $N=\overline{1,4}$ were constructed in explicit form.

In this work we are interested in constructing a so called vertex operator
representation \cite{Ne,Di,PV,Ve} for solutions to the  Gurevich-Zybin
hydrodynamical hierarchy \ (\ref{Z1}) at $N=2:$%
\begin{equation}
\left\{
\begin{array}{c}
D_{t}u=u_{t}+uu_{x}=v, \\
D_{t}v=v_{t}+uv_{x}=0,%
\end{array}%
\right.   \label{Z1aa}
\end{equation}%
making use an approach recently devised in \cite{BPP,BP} for the case of the
classical AKNS hierarchy of integrable flows, and which can be easily
generalized for treating the problem for arbitrary integers $N\in \mathbb{Z}%
_{+}.$

\section{\protect\bigskip A vertex operator analysis}

\bigskip We begin with a Lax type linear spectral problem \cite%
{Pav,GPPP,GBPPP} for the equation \ (\ref{Z1}) at $N=2:$%
\begin{equation}
\left\{
\begin{array}{c}
D_{t}u=u_{t}+uu_{x}=v, \\
D_{t}v=v_{t}+uv_{x}=0,%
\end{array}%
\right.  \label{Z1a}
\end{equation}%
defined on the space of smooth real-valued $2\pi $-periodic functions $%
(u,v)^{\intercal }\in M\subset C^{\infty }(\mathbb{R}/2\pi \mathbb{Z};%
\mathbb{R}^{2}):$%
\begin{equation}
df/dx=\text{\ }\ell \lbrack u,v;\lambda ]f,\text{ \ \ \ \ }\ell \lbrack
u,v;\lambda ]:=\left(
\begin{array}{cc}
-\lambda u_{x}/2 & -v_{x} \\
\lambda ^{2}/2 & \lambda u_{x}/2%
\end{array}%
\right) ,  \label{Z2}
\end{equation}%
where, by definition, $v:=D_{t}u,$ $f\in L_{\infty }(\mathbb{R}/2\pi \mathbb{%
Z};\mathbb{C}^{2})$ and $\lambda \in \mathbb{C}$ is a spectral parameter.
Assume that a vector function $(u,v)^{\top }\in M$ depends parametrically on
the infinite set $t:=\{t_{1},t_{2},t_{3},\ldots \}\in \mathbb{R}^{\mathbb{Z}%
_{+}}$ in such a way that the generalized Floquet spectrum \cite{Nov,FT,HPP}
$\sigma (\ell ):=\{\lambda \in \mathbb{C}:\sup_{x\in \mathbb{R}%
}||f(x;\lambda )||_{\infty }<\infty \}$ of the linear problem (\ref{Z2})
persists in being parametrically iso-spectral, that is $d\sigma (\ell
)/dt_{j}=0$ for all $t_{j}\in \mathbb{R}.$ The iso-spectrality condition
gives rise to a hierarchy of commuting to each other nonlinear
bi-Hamiltonian dynamical systems on the functional manifold $M$ in the
general form
\begin{equation}
\frac{d}{dt_{j}}(u(t),v(t))^{\top }=-\vartheta  {grad}%
H_{j}[u,v]:=K_{j}[u(t),v(t)],  \label{Z3}
\end{equation}%
where $K_{j}:M\rightarrow T(M)$ and $\ H_{j}\in \mathcal{D}(M),j\in \mathbb{Z%
}_{+},$ are, respectively, vector fields and conservation laws on the
manifold $M,$ which were before described in \cite{GPPP,GBPPP,PoP},
\begin{equation}
\vartheta :=\left(
\begin{array}{cc}
0 & \partial \\
\partial & 0%
\end{array}%
\right)  \label{Z4}
\end{equation}%
is a Poisson structure on the manifold $M$ and, by definition, \
\begin{equation}
\binom{u(t)}{v(t)}:=\binom{u(x,t_{1},t_{2,}t_{3,}...)}{%
v(x.t_{1},t_{2,}t_{3,}...)}  \label{Z5}
\end{equation}%
for $t\in \mathbb{R}^{\mathbb{N}}.$

\bigskip It is well known \cite{FT,HPP,Nov,PM} that the Casimir invariants,
determining conservation laws for dynamical systems (\ref{Z3}), are
generated by the suitably normalized monodromy matrix $\tilde{S}(x;\lambda
)\in End$ $\mathbb{C}^{2}$ of the linear problem \ (\ref{Z2})
\begin{equation}
\tilde{S}(x;\lambda )=k(\lambda )S(x;\lambda )-\frac{k(\lambda )}{2}\mathrm{%
tr}S(x;\lambda ),  \label{Z6}
\end{equation}%
where $F(y,x;\lambda )\in End$\thinspace $\ \mathbb{C}^{2}$ is the matrix
solution to the Cauchy problems%
\begin{equation}
\frac{d}{dy}F(y,x;\lambda )=\ell (y;\lambda )F(y,x;\lambda ),\text{ \ \ }%
F(y,x;\lambda )|_{y=x}=\mathbf{I,}  \label{Z7}
\end{equation}%
for all $\lambda \in \mathbb{C}\ $and $x,y\in \mathbb{R},$ where $\mathbf{%
I\in }End$ $\mathbb{C}^{2}$ is the identity matrix, $S(x;\lambda ):=F(x+2\pi
,x;\lambda )$ is the usual monodromy matrix for the equation \ (\ref{Z7}).
Here the parameter $k(\lambda )\in \mathbb{C}$ is invariant with respect to
flows \ (\ref{Z3}) and is chosen in such a way that the asymptotic condition
\begin{equation}
\tilde{S}(x;\lambda )\in \tilde{\mathcal{G}}_{-}  \label{Z8}
\end{equation}%
as $\lambda \rightarrow \infty $ holds for all $x\in \mathbb{R}.$ Here \ \ \
\ \ $\tilde{\mathcal{G}}_{-}\subset \mathcal{\tilde{G}},$ where $\mathcal{%
\tilde{G}}:=\tilde{\mathcal{G}}_{+}\oplus \tilde{\mathcal{G}}_{-}$ is the
natural splitting into two affine subalgebras of positive and negative $%
\lambda $-expansions of the centrally extended \cite{FT,RS-T} affine current
$\mathfrak{sl}(2)$-algebra \ $\hat{\mathcal{G}}:=\tilde{\mathcal{G}}\oplus
\mathbb{C}:$
\begin{equation}
\tilde{\mathcal{G}}:=\{a=\sum_{j\in \mathbb{Z},\,j\ll \infty }a^{(j)}\otimes
\lambda ^{j}:a^{(j)}\in C^{\infty }\left( \mathbb{R}/2\pi \mathbb{Z};%
\mathfrak{sl}(2;\mathbb{C})\right) \mathbb{\}}.  \label{Z9}
\end{equation}%
The latter is endowed with the Lie commutator
\begin{equation}
\lbrack (a_{1},c_{1}),(a_{2},c_{2})]:=([a_{1},a_{2}],\left\langle
a_{1},da_{2}/dx\right\rangle ),  \label{Z10}
\end{equation}%
where the scalar product is defined as
\begin{equation}
\left\langle a_{1},a_{2}\right\rangle :=\mathrm{res}_{\lambda =\infty
}\int_{0}^{2\pi }\mathrm{tr}(a_{1}a_{2})dx  \label{Z11}
\end{equation}%
for any two elements $a_{1},a_{2}\in \tilde{\mathcal{G}}$ \ \ with "$\mathrm{%
res"}$ and "$\mathrm{tr"}$ being the usual residue and trace maps,
respectively. As the spectrum $\sigma (\ell )\subset \mathbb{C}$ of the
problem \ (\ref{Z2}) is supposed to be parametrically independent, flows \ (%
\ref{Z3}) are naturally associated with evolution equations
\begin{equation}
d\tilde{S}/dt_{j}=[(\lambda ^{j+1}\tilde{S})_{+},\tilde{S}]  \label{Z11a}
\end{equation}%
for all $j\in \mathbb{R},$ which are generated by the set $I(\hat{\mathcal{G}%
^{\ast }})$ of Casimir invariants of the coadjoint action of the current
algebra \ $\hat{\mathcal{G}}$ \ on a given element $\ell (x;\lambda )\in
\tilde{\mathcal{G}}_{-}^{\ast }\cong \tilde{\mathcal{G}}_{+}$ contained in
the space of smooth functionals $\mathcal{D}(\hat{\mathcal{G}}).$ In
particular, a functional $\gamma (\lambda )\in I(\hat{\mathcal{G}})$ if and
only if
\begin{equation}
\lbrack \tilde{S}(x;\lambda ),\ell (x;\lambda )]+\frac{d}{dx}\tilde{S}%
(x;\lambda )=0,  \label{Z12}
\end{equation}%
where the gradient $\tilde{S}(x;\lambda ):=\mathrm{grad}\gamma (\lambda
)(\ell )\in \tilde{\mathcal{G}}_{-}$ is defined with respect to the scalar
product (\ref{Z11}) by means of the variation
\begin{equation}
\delta \gamma (\lambda ):=\left\langle \mathrm{grad}\gamma (\lambda )(\ell
),\delta \ell \right\rangle .  \label{Z13}
\end{equation}

To construct the solution to matrix equation (\ref{Z12}), we find
preliminary a partial solution $\tilde{F}(y,x;\lambda )\in End$\thinspace\ $%
\mathbb{C}^{2},$ $x,y\in \mathbb{R},$ to equation \ (\ref{Z7}) satisfying \
the asymptotic Cauchy data
\begin{equation}
\tilde{F}(y,x;\lambda )|_{y=x}=\mathbf{I}+O(1/\lambda )  \label{Z14}
\end{equation}%
as $\lambda \rightarrow \infty .$ It is easy to check that
\begin{equation}
\tilde{F}(y,x;\lambda )=\left(
\begin{array}{cc}
\tilde{e}_{1}(y,x;\lambda ) & -\frac{\tilde{\beta}(y;\lambda )}{\lambda }%
\tilde{e}_{2}(y,x;\lambda ) \\
-\frac{\lambda }{\tilde{\alpha}(y;\lambda )}\tilde{e}_{1}(y,x;\lambda ) &
\tilde{e}_{2}(y,x;\lambda )%
\end{array}%
\right) ,  \label{Z15}
\end{equation}%
is an exact functional solution to (\ref{Z7}) satisfying condition (\ref{Z14}%
), where we have defined
\begin{align}
\tilde{e}_{1}(y,x;\lambda )& :=\exp \{\frac{\lambda }{2}[u(x)-u(y)]+\lambda
\int_{x}^{y}\tilde{\alpha}\text{ }dv(s)\},  \label{Z16} \\
\tilde{e}_{2}(y,x;\lambda )& :=\exp \{\frac{\lambda }{2}[u(y)-u(x)]-\frac{%
\lambda }{2}\int_{x}^{y}\tilde{\beta}\text{ }ds\},  \notag
\end{align}%
with the vector-functions $\alpha ^{\pm }\in C^{\infty }(\mathbb{R}/2\pi
\mathbb{Z};\mathbb{R})$ satisfying the following determining functional
relationships:%
\begin{eqnarray}
\tilde{\alpha} &=&u_{x}+(u_{x}^{2}-2v_{x}+\xi \tilde{\alpha})^{1/2},\text{ \
}  \notag \\
\tilde{\beta} &=&u_{x}-(u_{x}^{2}-2v_{x}+\xi \tilde{\beta})^{1/2},
\label{Z17}
\end{eqnarray}%
as $\xi :=1/\lambda \rightarrow 0$ and existing when the condition $\varphi
(x,t):=\sqrt{u_{x}^{2}-2v_{x}}\neq 0$ on the manifold $M$ at $t=0\in \mathbb{%
R}^{\mathbb{N}}.$

The fundamental matrix $F(y,x;\lambda )\in End$\thinspace\ $\mathbb{C}^{2}$
can be represented for all $x,y\in \mathbb{R}$ in the form%
\begin{equation}
F(y,x;\lambda )=\tilde{F}(y,x;\lambda )\tilde{F}^{-1}(x,x;\lambda ).
\label{Z18}
\end{equation}%
Consequently, if one sets $y=x+2\pi $ in this formula and defines the
expression
\begin{equation}
k(\lambda ):=\lambda ^{-1}[\tilde{e}_{1}(x+2\pi ,x;\lambda )-\tilde{e}%
_{2}(x+2\pi ,x;\lambda )]^{-1},  \label{Z19}
\end{equation}%
it follows from (\ref{Z6}), \ (\ref{Z15}) and (\ref{Z18}) that the exact
functional matrix representation \

\begin{equation}
\tilde{S}(x;\lambda )=\left(
\begin{array}{cc}
\frac{\lbrack \tilde{\alpha}(x;\lambda )+\tilde{\beta}(x;\lambda )]}{%
2\lambda \lbrack \tilde{\alpha}(x;\lambda )-\tilde{\beta}(x;\lambda )]} &
\frac{\tilde{\alpha}\tilde{\beta}}{\lambda ^{2}[\tilde{\alpha}(x;\lambda )-%
\tilde{\beta}(x;\lambda )]} \\
-\frac{1}{[\tilde{\alpha}(x;\lambda )-\tilde{\beta}(x;\lambda )]} & \frac{[%
\tilde{\beta}(x;\lambda )+\tilde{\alpha}(x;\lambda )]}{2\lambda \lbrack
\tilde{\beta}(x;\lambda )-\tilde{\alpha}(x;\lambda )]}%
\end{array}%
\right) ,  \label{Z20}
\end{equation}%
satisfies the necessary condition (\ref{Z8}) as $\lambda \rightarrow \infty
. $

\begin{remark}
\label{Rm_Z1}The invariance of the expression (\ref{Z19}) with respect to
the generating vector field (\ref{Z3}) on the manifold $M$ derives from the
representation (\ref{Z18}), the equations (\ref{Z12}) and
\begin{equation}
\frac{d}{dt}\tilde{F}(y,x_{0};\mu )=\frac{\lambda ^{3}}{\mu -\lambda }\tilde{%
S}(x;\lambda )\tilde{F}(y,x_{0};\mu ),  \label{Z21}
\end{equation}%
which follows naturally from the determining matrix flows (\ref{Z11a}) upon
applying the translation $y\rightarrow y+2\pi .$
\end{remark}

The matrix expression (\ref{Z20}) gives rise to the following important
functional relationships:%
\begin{equation}
\frac{1-\lambda (\tilde{s}_{11}-\tilde{s}_{22})}{2\tilde{s}_{21}}=\tilde{%
\alpha},\text{ }\frac{-2\lambda ^{2}\tilde{s}_{12}}{1-\lambda (\tilde{s}%
_{11}-\tilde{s}_{22})}=\tilde{\beta},  \label{Z22}
\end{equation}%
which allow to introduce in a natural way the vertex operator vector fields%
\begin{equation}
\text{ \ \ }X_{\lambda }^{\pm }=\exp (\pm D_{\lambda }),\text{ \ \ }%
D_{\lambda }:=\sum_{j\in \mathbb{Z}_{+}}\frac{1}{(j+1)}\lambda ^{-(j+1)}%
\frac{d}{dt_{j+1}},\text{ \ \ \ \ \ }  \label{Z23}
\end{equation}%
acting on an arbitrary smooth function $\eta \in C^{\infty }(\mathbb{R}^{%
\mathbb{Z}_{+}};\mathbb{R})$ by means of the shifting mappings:%
\begin{equation}
\begin{array}{c}
X_{\lambda }^{\pm }\text{ }\eta (x,t_{1},t_{2},...,t_{j},...):=\eta ^{\pm
}(x,t;\lambda )= \\
=\eta (x,t_{1}\pm 1/\lambda ,t_{2}\pm /(2\lambda ^{2}),t_{3}\pm 1/(3\lambda
^{3})...,t_{j}\pm 1/(j\lambda ^{j}),...)%
\end{array}
\label{Z24}
\end{equation}%
as $\lambda \rightarrow \infty .$ Namely, we following proposition holds.

\begin{proposition}
\bigskip The \ functional vertex operator expressions%
\begin{eqnarray}
\tilde{\alpha}(x,t;\lambda ) &=&X_{\lambda }^{-}\alpha (x,t)=\alpha
^{-}(x,t;\lambda ),  \label{Z25} \\
\tilde{\beta}(x,t;\lambda ) &=&X_{\lambda }^{+}\beta (x,t))=\beta
^{+}(x,t;\lambda )  \notag
\end{eqnarray}%
solve the functional equations \ (\ref{Z17}), that is%
\begin{eqnarray}
\alpha ^{-} &=&u_{x}+(u_{x}^{2}-2v_{x}+\xi \alpha ^{-})^{1/2},\text{ \ }
\notag \\
\beta ^{+} &=&u_{x}-(u_{x}^{2}-2v_{x}+\xi \beta ^{+})^{1/2},  \label{Z25a}
\end{eqnarray}%
where $t\in \mathbb{R}^{\mathbb{Z}_{+}}$and $\xi =1/\lambda \rightarrow 0.$
\end{proposition}

\begin{proof}
To state this proposition it is enough to show that the following
relationships hold:
\begin{eqnarray}
\frac{d}{d\xi }\left[ \frac{1-\lambda (\tilde{s}_{11}-\tilde{s}_{22})}{2%
\tilde{s}_{21}}\right] _{\lambda =1/\xi } &=&\frac{d}{dt}\left[ \frac{%
1-\lambda (\tilde{s}_{11}-\tilde{s}_{22})}{2\tilde{s}_{21}}\right] _{\lambda
=1/\xi },  \notag \\
\frac{d}{d\xi }\left[ \frac{-8\lambda ^{2}\tilde{s}_{12}}{1-\lambda (\tilde{s%
}_{11}-\tilde{s}_{22})}\right] _{\lambda =1/\xi } &=&\frac{d}{dt}\left[
\frac{-8\lambda ^{2}\tilde{s}_{12}}{1-\lambda (\tilde{s}_{11}-\tilde{s}_{22})%
}\right] _{\lambda =1/\xi }  \label{Z26}
\end{eqnarray}%
for any parameter $\xi \rightarrow 0,$ where by definition%
\begin{equation}
\frac{d}{dt}:=\left. \frac{d}{d\xi }D_{\lambda }\right\vert _{\lambda =1/\xi
}=\sum_{j\in \mathbb{Z}_{+}}\xi ^{j}\frac{d}{dt_{j+1}}  \label{Z26a}
\end{equation}%
is a generating evolution vector field. \ Before doing this we find the
evolution equation
\begin{equation}
\frac{d}{dt}\tilde{S}(x;\mu )=[\lambda ^{3}\frac{d}{d\lambda }\tilde{S}%
(x;\mu ),\tilde{S}(x;\lambda )]  \label{Z27}
\end{equation}%
on the matrix $\tilde{S}(x;\mu )$ as $\ \mu ,\lambda \rightarrow \infty ,$
which entails the following differential relationships:
\begin{equation}
\begin{array}{c}
d\tilde{s}_{11}/dt=\lambda ^{3}(\tilde{s}_{21}d\tilde{s}_{12}/d\lambda -%
\tilde{s}_{12}d\tilde{s}_{21}/d\lambda ), \\
d\tilde{s}_{22}/dt=\lambda ^{3}(\tilde{s}_{12}d\tilde{s}_{21}/d\lambda -%
\tilde{s}_{21}d\tilde{s}_{12}/d\lambda ), \\
d\tilde{s}_{22}/dt=\lambda ^{3}[\tilde{s}_{12}\frac{d}{d\lambda }(\tilde{s}%
_{11}-\tilde{s}_{22})-(\tilde{s}_{11}-\tilde{s}_{22})\frac{d\tilde{s}_{12}}{%
d\lambda }), \\
d\tilde{s}_{11}/dt=\lambda ^{3}[\tilde{s}_{21}\frac{d}{d\lambda }(\tilde{s}%
_{22}-\tilde{s}_{11})-(\tilde{s}_{22}-\tilde{s}_{11})\frac{d\tilde{s}_{21}}{%
d\lambda }).%
\end{array}
\label{Z28}
\end{equation}%
Using these relationships (\ref{Z28}), one can easily obtain by means of
simple, but rather cumbersome calculations, the needed relationships \ (\ref%
{Z26}). As their direct consequences the \textit{vertex operator
representations }\ (\ref{Z25}) for the vector functions $\tilde{\alpha},%
\tilde{\beta}\in C(\mathbb{R}^{\mathbb{Z}_{+}};\mathbb{R})$ hold.
\end{proof}

Now we take into account that, owing to the determining functional
representations (\ref{Z17}), that the limits$^{\infty }$
\begin{eqnarray}
\lim_{\lambda \rightarrow \infty }\alpha ^{-}(x,t;\lambda )
&=&u_{x}(x,t)+\varphi (x,t),\text{ \ \ }  \label{Z29} \\
\lim_{\lambda \rightarrow \infty }\beta ^{+}(x,t;\lambda )
&=&u_{x}(x,t)-\varphi (x,t),\text{~\ }\varphi (x,t):=\sqrt{%
u_{x}^{2}(x,t)-2v_{x}(x,t)},  \notag
\end{eqnarray}%
exist on the manifold $M.$ \ Moreover, having iterated the functional
relationships \ (\ref{Z17}), one can find that
\begin{align}
X_{\lambda }^{-}\alpha & =\alpha ^{-}=u_{x}+\varphi +\xi (\frac{u_{xx}}{%
\varphi }+\frac{\varphi _{x}}{\varphi })+  \notag \\
& +\frac{\xi ^{2}}{2}(\frac{u_{xx}^{2}+2u_{xx}\varphi _{x}-u_{3x}\varphi }{%
\varphi ^{3}}+\frac{\varphi _{xx}\varphi +5\varphi _{x}^{2}}{\varphi ^{3}}%
)+O(\xi ^{3}),  \notag \\
X_{\lambda }^{+}\beta & =\beta ^{+}=u_{x}-\varphi -\xi (\frac{u_{xx}}{%
\varphi }-\frac{\varphi _{x}}{\varphi })-  \label{Z30} \\
& -\frac{\xi ^{2}}{2}(\frac{u_{xx}^{2}-2u_{xx}\varphi _{x}+u_{3x}\varphi }{%
\varphi ^{3}}+\frac{\varphi _{xx}\varphi +5\varphi _{x}^{2}}{\varphi ^{3}}%
)+O(\xi ^{3}),  \notag
\end{align}%
which immediately yield the higher Riemann type commuting nonlinear Lax
integrable dispersive dynamical systems on the functional manifold $M.$ For
instance, making use of the relationships
\begin{equation}
\lim_{\lambda \rightarrow \infty }[\alpha ^{-}(x,t;\lambda )\pm \beta
^{+}(x,t;\lambda )]/2=\left\{
\begin{array}{c}
u_{x}(x,t), \\
\varphi (x,t)%
\end{array}%
\right. ,  \label{Z30a}
\end{equation}%
one easily obtains that
\begin{equation}
\frac{d}{dt_{1}}\binom{u_{x}}{\varphi }=\binom{-u_{xx}/\varphi }{-\varphi
_{x}/\varphi },\frac{d}{dt_{2}}\binom{u_{x}}{\varphi }=\binom{%
(u_{xx}^{2}+7\varphi _{x}^{2})/\varphi ^{3}}{(2u_{3x}\varphi -4u_{x}\varphi
_{x})/\varphi ^{3}},...,  \label{Z31}
\end{equation}%
and so on, where \ $\varphi =\sqrt{u_{x}^{2}-2v_{x}}$ and we took into
account that the following asymptotic expansions hold%
\begin{equation}
\begin{array}{c}
X_{\lambda }^{-}\alpha (x,t;\lambda )=u_{x}+\varphi -\xi
(u_{x,t_{1}}+\varphi _{t_{1}})+ \\
+\frac{\xi ^{2}}{2}(u_{x,t_{1},t_{1}}+\varphi
_{t_{1},t_{1}}-u_{x,t_{2}}-\varphi _{t_{2}})+O(\xi ^{3}), \\
X_{\lambda }^{+}\beta (x,t;\lambda )=u_{x}-\varphi +\xi (u_{x,t_{1}}-\varphi
_{t_{1}})+ \\
+\frac{\xi ^{2}}{2}(u_{x,t_{1},t_{1}}-\varphi
_{t_{1},t_{1}}+u_{x,t_{2}}-\varphi _{t_{2}})+O(\xi ^{3})%
\end{array}
\label{Z32}
\end{equation}%
as $\xi =1/\lambda \rightarrow 0.$

It is worth here to mention\ that the scheme devised above for finding the
corresponding vertex operator representations for the Riemann type equation
\ (\ref{Z1a}) can be similarly generalized for treating others equations of
the infinite \ hierarchy \ (\ref{Z1}) when $N\geq 3,$ having taking into
account the existence of their suitable Lax type representations found before
in recent works \cite{PAPP,GPPP,GBPPP}.

\section{ Concluding remarks}

The vertex operator functional representations of the solution to the
Riemann type hydrodynamical equation (\ref{Z1a}) in the form \ (\ref{Z25a})
is crucially based on the representations (\ref{Z22}) and evolution
equations (\ref{Z26}), which provide a very straightforward and transparent
explanation of many of \textquotedblleft miraculous\textquotedblright\
vertex operator calculations presented before both in \cite{Ne,Di} and in
\cite{PV}. It should be noted that the effectiveness of our approach to
studying the vertex operator representation of the Riemann type hierarchy
owes much to the important exact representation (\ref{Z20}) for the
corresponding monodromy matrix, whose properties are described by means of
applying the standard \ \cite{FT,HPP,PM,Bl} Lie-algebraic techniques. As an
indication of possible future research, it should also be mentioned that it
would be interesting to generalize the vertex operator approach devised in
this work to other linear spectral problems such as those related to
dynamical systems with a parametrical spectral \cite{BZM,BPS,HPP}
dependence, spatially two-dimensional \cite{ZM}, Pavlov's and heavenly \cite%
{MS} dynamical systems.

\section{Acknowledgments}

D. Blackmore wishes to thank the National Science Foundation for support
from NSF Grant CMMI - 1029809 and his coauthor for enlisting him in the
efforts that produced this paper. A.K. Prykarpatsky cordially thanks Profs.
N. Bogolubov (Jr.) for useful discussions of the results obtained.

\end{document}